\documentclass[a4paper,onecolumn]{article}
\usepackage{amsfonts}
\usepackage{graphicx}
\usepackage{amsmath}

\input{tcilatex}
\begin{document}

\title{About the Bidimensional Beer-Lambert Law }
\author{B. Lacaze \\
%EndAName
Tesa 14/16 Port St-Etienne 31000 Toulouse France\\
e-mail address: bernard.lacaze@tesa.prd.fr}
\maketitle

\begin{abstract}
In acoustics, ultrasonics and in electromagnetic wave propagation, the
crossed medium can be often modelled by a linear invariant filter (LIF)
which acts on a wide-sense stationary process. Its complex gain follows the
Beer-Lambert law i.e is in the form $\exp \left[ -\alpha z\right] $ where $z$
is the thickness of the medium and $\alpha $ depends on the frequency and on
the medium properties. This paper addresses a generalization for
electromagnetic waves when the beam polarization has to be taken into
account. In this case, we have to study the evolution of both components of
the electric field (assumed orthogonal to the trajectory). We assume that
each component at $z$ is a linear function of both components at 0. New
results are obtained modelling each piece of medium by four LIF. They lead
to a great choice of possibilities in the medium modelling. Particular cases
can be deduced from works of R. C. Jones on deterministic monochromatic
light.

\textit{keywords: }linear filtering, polarization, Beer-Lambert law, random
processes.
\end{abstract}

\section{Introduction}

\subsection{The Beer-Lambert law}

The Beer-Lambert law (B.L law) states that some positive quantity $A\left(
0\right) $ at the input of some medium varies following the equation 
\begin{equation}
A\left( z\right) =A\left( 0\right) e^{-\alpha z}
\end{equation}%
where $z$ is the covered distance and $\alpha $ is a parameter defined by
the medium \cite{Ishi}. The equality $\left( 1\right) $ comes from the
approximation 
\begin{equation*}
A\left( z+dz\right) -A\left( z\right) \approx -\alpha A\left( z\right) dz
\end{equation*}%
which postulates that the evolution of $A\left( z\right) $ on a small
thickness $dz$ of the medium is proportional to $A\left( z\right) $ and to $%
dz$, with a coefficient $\alpha >0$ which is defined by the medium. Then the
differential equation $A^{\prime }\left( z\right) =-\alpha A\left( z\right) $
which leads to $\left( 1\right) .$ A more general method starts from the
equality%
\begin{equation}
A\left( 0\right) A\left( z+z^{\prime }\right) =A\left( z\right) A\left(
z^{\prime }\right)
\end{equation}%
whatever $z,z^{\prime }.$ Equivalently the quotient $A\left( z+z^{\prime
}\right) /A\left( z\right) $ depends only on $z^{\prime },$ and any piece of
the medium of length $z$ has the same behavior. If we assume that $A\left(
z\right) $ is a continuous function on $\mathbb{R}^{+}$, the only solution
of $\left( 2\right) $ is $\left( 1\right) $ for some $\alpha \in \mathbb{C}.$

If we take $A\left( 0\right) =e^{i\omega _{0}t}$, $\left( 1\right) $ becomes
($\mathcal{R}\left[ \alpha \right] $ and $\mathcal{I}\left[ \alpha \right] $
stand for the real and imaginary parts of $\alpha )$ 
\begin{equation}
A\left( z\right) =\exp \left[ i\omega _{0}\left( t-\frac{z}{\omega _{0}}%
\mathcal{I}\left[ \alpha \right] \right) -z\mathcal{R}\left[ \alpha \right] %
\right] .
\end{equation}%
This means that the monochromatic wave $e^{i\omega _{0}t}$ is delayed by $%
\frac{z}{\omega _{0}}\mathcal{I}\left[ \alpha \right] $ and weakened by exp$%
\left[ -z\mathcal{R}\left[ \alpha \right] \right] $ when crossing a
thickness $z$ in the medium$.$

Now we place ourselves from a signal processing perspective. We assume that
a piece of medium of any thickness $z$ is equivalent to a LIF (Linear
Invariant Filter) $\mathcal{F}_{z}$ with complex gain $F_{z}\left( \omega
\right) $ and that any piece of thickness $z+z^{\prime }$ has the behavior
of two filters in series $\mathcal{F}_{z}$ and $\mathcal{F}_{z^{\prime }}$ 
\cite{Papo}, \cite{Laca1}$.$ This means that whatever the frequency $\omega
/2\pi $ 
\begin{equation}
F_{z+z^{\prime }}\left( \omega \right) =F_{z}\left( \omega \right)
F_{z^{\prime }}\left( \omega \right) .
\end{equation}%
Obviously we take $F_{0}=1.$ What preceeds implies that for each $\omega $
it exists a complex $\alpha \left( \omega \right) $ such that%
\begin{equation*}
F_{z}\left( \omega \right) =e^{-z\alpha \left( \omega \right) }
\end{equation*}%
By definition $F_{z}\left( \omega \right) e^{i\omega t}$ is the output of
the filter $\mathcal{F}_{z}$ when the input is $e^{i\omega t}.$ For such an
input the power $P_{z}$ at the output is 
\begin{equation}
P_{z}=e^{-2z\mathcal{R}\left[ \alpha \left( \omega \right) \right] }
\end{equation}%
$\left( 5\right) $ summarizes the Beer-Lambert law for wave propagation
through a continuous medium, used in acoustics, ultrasonics and
electromagnetics. $\alpha \left( \omega \right) $ gives the attenuation of
the wave (by its real part) and the celerity of the wave (by its imaginary
part). The Kramers-Kronig relation links the real and imaginary parts of $%
F_{z}\left( \omega \right) $ which constitute a pair of Hilbert transforms 
\cite{Wate}, \cite{Laca3}. $\left( 5\right) $ is matched to monochromatic
waves. More generally when the stationary process $\mathbf{Z}_{0}=\left\{
Z_{0}\left( t\right) ,t\in \mathbb{R}\right\} $ is the input of $\mathcal{F}%
_{z}$, the output $\mathbf{Z}_{z}=\left\{ Z_{z}\left( t\right) ,t\in \mathbb{%
R}\right\} $ verifies (E$\left[ ..\right] $ stands for mathematical
expectation or ensemble mean)%
\begin{equation*}
P_{z}=\text{E}\left[ \left\vert Z_{z}\left( t\right) \right\vert ^{2}\right]
=\int_{-\infty }^{\infty }e^{-2z\mathcal{R}\left[ \alpha \left( \omega
\right) \right] }s_{0}\left( \omega \right) d\omega
\end{equation*}%
where $s_{0}\left( \omega \right) $ is the power spectral density of $%
\mathbf{Z}_{0}$ ($s_{0}\left( \omega \right) =\delta \left( \omega -\omega
_{0}\right) $ for a unit power monochromatic wave at $\omega _{0}).$ The
fact that measurements are generally performed by non-monochromatic waves
lead to gaps with the Beer-Lambert law in the form $\left( 6\right) $ \cite%
{Alle}. Also, B.L law can be untrue when powers are too high (which act on
the medium properties) \cite{Abit} or when the beam expands \cite{Laca2}.
However, B.L law has many applications in physics, chemistry and medecine 
\cite{Hodg}, \cite{Klin}, \cite{Fish}, \cite{Lewi}.

\subsection{A counter-example}

The Beer-Lambert law applies for light amplitude or power after crossing
continuous permanent media which can be viewed as a set of filters in
series. However light or radar wave is not only defined by an amplitude \cite%
{Born}, \cite{Mand}. Electric field \textbf{E}$^{z}=\left( \mathbf{E}%
_{x}^{z},\mathbf{E}_{y}^{z}\right) $ is a two-dimensional vector with
respect to axes Ox and Oy at distance $z$ of origin O and with components 
\begin{equation*}
\mathbf{E}_{x}^{z}=\left\{ E_{x}^{z}\left( t\right) ,t\in \mathbb{R}\right\}
,\mathbf{E}_{y}^{z}=\left\{ E_{y}^{z}\left( t\right) ,t\in \mathbb{R}%
\right\} .
\end{equation*}%
It is assumed that the beam propagates in the neighbourood of the axis Oz
and that the field is orthogonal to this axis. The beam is polarized in the
direction $\theta $ at $z$ when%
\begin{equation*}
\left\{ 
\begin{array}{c}
E_{x}^{z}\left( t\right) =A\left( t\right) \cos \theta \\ 
E_{y}^{z}\left( t\right) =A\left( t\right) \sin \theta%
\end{array}%
\right.
\end{equation*}%
for some (real or complex) process $\mathbf{A=}\left\{ A\left( t\right)
,t\in \mathbb{R}\right\} .$ The beam is unpolarized at $z$ when the
cross-spectrum $s_{xy}^{z}\left( \omega \right) $ of components verifies%
\begin{equation*}
s_{xy}^{z}\left( \omega \right) =0
\end{equation*}%
whatever the basis Oxy. This implies the equality of power spectra. The wave
is partially polarized in other cases.

The power $P_{z}$ at $z$ for a stationary wave is defined by%
\begin{equation}
P_{z}=\text{E}\left[ \left\vert E_{x}^{z}\left( t\right) \right\vert ^{2}%
\right] +\text{E}\left[ \left\vert E_{y}^{z}\left( t\right) \right\vert ^{2}%
\right] .
\end{equation}%
We know that media act upon polarization and then can influence
measurements, for instance in the case of antennas (generally matched to a
particular polarization state). B.L law is not available for behavior of a
given component of the electric field except particular cases. For instance,
take the wave%
\begin{equation*}
E_{x}^{0}\left( t\right) =e^{i\omega _{0}t},\text{ \ }E_{y}^{0}\left(
t\right) =0
\end{equation*}%
which propagates in a medium which rotates the beam by angle proportional to
thickness $z.$ We have ($c$ is the celerity in the medium)%
\begin{equation*}
E_{x}^{z}\left( t\right) =e^{i\omega _{0}(t-z/c)}\cos \left[ z\alpha \left(
\omega _{0}\right) \right]
\end{equation*}%
If we measure $E_{x}^{z}\left( t\right) $ the medium is equivalent to a
filter of complex gain 
\begin{equation*}
F_{z}\left( \omega \right) =e^{-i\omega z/c(\omega )}\cos \left[ z\alpha
\left( \omega \right) \right]
\end{equation*}%
when a particular direction is chosen ($F_{z}\left( \omega \right)
=e^{-i\omega z/c}\sin \left[ z\alpha \left( \omega \right) \right] $ for the
orthogonal direction and $c$ can depend on $\omega $). The term $\alpha
\left( \omega \right) $ takes into account a possible dependency of the
rotation angle with the frequency. For an antenna which selects the
component in a given direction at distance $z$, the B.L law is not verified.

\subsection{The two-dimensional case}

The B.L law is established for one-component monochromatic waves (i.e for
waves defined by only one quantity depending on the time $t$ and on the
space coordinate $z)$. They are time-functions in the form $ae^{i\omega
_{0}t}$ ($a\in \mathbb{C}$ depends on $z$). Quasi-monochromatic waves can be
defined as random processes in the form%
\begin{equation*}
E^{z}\left( t\right) =A^{z}\left( t\right) e^{i\omega _{0}t}
\end{equation*}%
where $\mathbf{A}^{z}=\left\{ A^{z}\left( t\right) ,t\in \mathbb{R}\right\} $
is stationary with a baseband spectrum which cancels outside $\left( -\omega
_{1},\omega _{1}\right) $ with $\omega _{1}/\omega _{0}\ll 1.$ The spectral
band of \textbf{E}$^{z}$ is included in the interval $\left( \omega
_{0}-\omega _{1},\omega _{0}+\omega _{1}\right) .$

In the two-components case (i.e for waves defined by two quantities $%
E_{x}^{z}\left( t\right) ,E_{y}^{z}\left( t\right) $ depending on the time $%
t $ and on the space coordinate $z)$ a quasi-monochromatic beam is defined
by 
\begin{equation}
E_{x}^{z}\left( t\right) =e^{i\omega _{0}t}A_{x}^{z}\left( t\right) ,\text{
\ }E_{y}^{z}\left( t\right) =e^{i\omega _{0}t}A_{y}^{z}\left( t\right)
\end{equation}%
where $\mathbf{A}_{x}^{z},\mathbf{A}_{y}^{z}$ are stationary with stationary
correlation and power spectrum inside $\left( -\omega _{1},\omega
_{1}\right) .$ These properties remain whatever the chosen coordinates axes.
It is reasonable to say that a wave is (purely) monochromatic when $\mathbf{A%
}_{x}^{z},\mathbf{A}_{y}^{z}$ verify a relation in the form (for some
constant $k$ and real $B_{z}\left( t\right) $) 
\begin{equation*}
A_{x}^{z}\left( t\right) =k\cos B^{z}\left( t\right) ,\ A_{y}^{z}\left(
t\right) =k\sin B^{z}\left( t\right)
\end{equation*}%
$B_{z}\left( t\right) $ represents the orientation and $ke^{i\omega _{0}t}$
the complex amplitude of the field. This means that its (complex) amplitude
is purely monochromatic in the usual sense. When $\mathbf{B}^{z}$ is
degenerate ($B^{z}\left( t\right) $ does not depend on $t$), the wave is
polarized (at $z).$ In other cases, the wave is partially polarized. When $%
\mathbf{E}_{x}^{z}$ and $\mathbf{E}_{y}^{z}$ have same power spectrum and
are uncorrelated, the wave is unpolarized at $z$. Both properties are true
in any orthogonal system.\ In optics, beams are often quasi-monochromatic
and partially polarized but it is not always true, particulary in astronomy
and communications. Polarized and unpolarized beams are convenient
idealizations.

The concept of quasi-monochromatic wave is often bad-fitted even in the
light domain. For instance Wolf-Rayet stars and B.L Lacertae have lines of
relative width larger than few percents. Idem for LEDs (light-emitting
diodes) with width larger than 10\%. A more general model has to be taken in
these cases.

We have seen that the B.L law for one-component beams can be proved using a
decomposition of media by LIF in series. The aim of this paper is to
generalize the B.L law in the two-components case, using signal theory and
modelling media as more general circuits. Inputs and outputs of these
circuits are stationary processes which represent the components of the
field. In the following section we consider that each component of the field
at a distance $z$ is the sum of two LIF outputs. The field components at the
origin point ($z=0)$ are the inputs of these LIF. This model allows to
determine the shape of the LIF characteristics, i.e the B.L law for
bi-dimensional beams.

Because two inputs and two outputs, 2x2 matrices of filters complex gains
will be defined. In the years 1940, R. C. Jones had developed a "New
Calculus for the Treatment of Optical Systems" \cite{Jone}. It was based on
2x2 matrices which act in the time domain on purely monochromatic waves.
Though formally results of Jones are very close to formulas of this paper,
they do not address the same objects. Actually Jones papers do not mention
Beer neither Lambert or Bouguer, the pioneer of this topic.

\section{Two-dimensional Beer-Lambert law}

We deal with a beam which propagates in direction Oz of the orthogonal
trihedron Oxyz. The electric field \textbf{E}$^{z}$ at time $t$ is defined
by its components $E_{x}^{z}\left( t\right) ,E_{y}^{z}\left( t\right) $ on
axes Ox and Oy at distance $z$ of the origin O. We assume that the medium
between $u$ and $u+z$ is defined by a set of 2x2 \textquotedblleft
scattering matrice\textquotedblright\ $\mathbf{H}^{z}$ independent of $u$%
\begin{equation*}
\mathbf{H}^{z}=\left[ 
\begin{array}{cc}
H_{11}^{z} & H_{12}^{z} \\ 
H_{21}^{z} & H_{22}^{z}%
\end{array}%
\right] 
\end{equation*}%
where the $H_{jk}^{z}\left( \omega \right) $ depend on the frequency $\omega
/2\pi $ and are complex gains of LIF $\mathcal{H}_{jk}^{z}$ such that%
\begin{equation}
\left\{ 
\begin{array}{c}
E_{x}^{u+z}\left( t\right) =\mathcal{H}_{11}^{z}\left[ \mathbf{E}_{x}^{u}%
\right] \left( t\right) +\mathcal{H}_{12}^{z}\left[ \mathbf{E}_{y}^{u}\right]
\left( t\right)  \\ 
E_{y}^{u+z}\left( t\right) =\mathcal{H}_{21}^{z}\left[ \mathbf{E}_{x}^{u}%
\right] \left( t\right) +\mathcal{H}_{22}^{z}\left[ \mathbf{E}_{y}^{u}\right]
\left( t\right) 
\end{array}%
\right. 
\end{equation}%
Equivalently the electric field \textbf{E}$^{z+u}$ at $z+u$ is linearly
dependent on its value \textbf{E}$^{u}$ at $u$ \cite{Laca4}$.$ The linearity
is expressed by the set of the LIF $\mathcal{H}_{jk}^{z}$ which depend only
on the medium. It is an obvious generalization of what was explained for the
one-component waves propagation. Filters parameters depend only on the
thickness of the considered medium. However results can be obtained without
this hypothesis. Figure 1 shows equivalent circuits of $\left( 8\right) .$ 

Filters in series lead to multiplication of complex gains and filters in
parallel to addition. The figure 2 summarizes the following equality%
\begin{equation}
\mathbf{H}^{z+u}=\mathbf{H}^{z}\mathbf{H}^{u}.
\end{equation}

$\left( 9\right) $ is equivalent to%
\begin{equation}
\left\{ 
\begin{array}{c}
H_{11}^{z+u}=H_{11}^{z}H_{11}^{u}+H_{12}^{z}H_{21}^{u} \\ 
H_{12}^{z+u}=H_{11}^{z}H_{12}^{u}+H_{12}^{z}H_{22}^{u} \\ 
H_{21}^{z+u}=H_{21}^{z}H_{11}^{u}+H_{22}^{z}H_{21}^{u} \\ 
H_{22}^{z+u}=H_{21}^{z}H_{12}^{u}+H_{22}^{z}H_{22}^{u}%
\end{array}%
\right.
\end{equation}%
We assume that the derivatives $h_{jk}^{0}=\frac{\partial }{\partial z}%
H_{jk}^{0}$ are finite. The first equation of $\left( 10\right) $ can be
written as%
\begin{equation*}
\frac{H_{11}^{z+u}-H_{11}^{z}}{u}=H_{11}^{z}\frac{H_{11}^{u}-1}{u}+H_{12}^{z}%
\frac{H_{21}^{u}}{u}
\end{equation*}%
Obviously we have $H_{11}^{0}=H_{22}^{0}=1$ and $H_{12}^{0}=H_{21}^{0}=0.$
When $u\rightarrow 0,$ we obtain (similar operation is done in the other
equations)%
\begin{equation}
\left\{ 
\begin{array}{c}
h_{11}^{z}=H_{11}^{z}h_{11}^{0}+H_{12}^{z}h_{21}^{0} \\ 
h_{12}^{z}=H_{11}^{z}h_{12}^{0}+H_{12}^{z}h_{22}^{0} \\ 
h_{21}^{z}=H_{21}^{z}h_{11}^{0}+H_{22}^{z}h_{21}^{0} \\ 
h_{22}^{z}=H_{21}^{z}h_{12}^{0}+H_{22}^{z}h_{22}^{0}.%
\end{array}%
\right.
\end{equation}%
The differential system can be split in two subsystems (equ.1+2 and
equ.3+4). We assume that%
\begin{equation}
\lim_{z\rightarrow \infty }H_{jk}^{z}=0
\end{equation}%
because any wave is evanescent in a passive medium. Two cases can be
highlighted following the (complex) eigenvalues $\lambda _{1},\lambda _{2}$
of\ the matrix%
\begin{equation*}
\left[ 
\begin{array}{cc}
h_{11}^{0} & h_{21}^{0} \\ 
h_{12}^{0} & h_{22}^{0}%
\end{array}%
\right] .
\end{equation*}%
Solutions are in the form following that $\lambda _{1}\neq \lambda _{2}$ or $%
\lambda _{1}=\lambda _{2}=\lambda $%
\begin{equation*}
\begin{array}{c}
H_{jk}^{z}=c_{jk1}e^{\lambda _{1}z}+c_{jk2}e^{\lambda _{2}z} \\ 
\text{or }H_{jk}^{z}=\left( c_{jk1}z+c_{jk2}\right) e^{\lambda z}.%
\end{array}%
\end{equation*}%
The eigenvalues cannot cancel (or corresponding coefficients cancel).
because $\left( 12\right) .$ Taking into account the initial conditions%
\begin{equation}
H_{11}^{0}=H_{22}^{0}=1\text{ and }H_{12}^{0}=H_{21}^{0}=0
\end{equation}%
leads to two cases

\textbf{Case\ 1: }$\lambda _{1}\neq \lambda _{2}$

\bigskip By identification with $\left( 11\right) $ we obtain%
\begin{equation}
\left\{ 
\begin{array}{l}
H_{11}^{z}=\alpha e^{\lambda _{1}z}+\left( 1-\alpha \right) e^{\lambda _{2}z}
\\ 
H_{12}^{z}=\frac{-h_{12}^{0}}{\lambda _{2}-\lambda _{1}}\left( e^{\lambda
_{1}z}-e^{\lambda _{2}z}\right) \\ 
H_{21}^{z}=\frac{-h_{21}^{0}}{\lambda _{2}-\lambda _{1}}\left( e^{\lambda
_{1}z}-e^{\lambda _{2}z}\right) \\ 
H_{22}^{z}=\left( 1-\alpha \right) e^{\lambda _{1}z}+\alpha e^{\lambda _{2}z}
\\ 
\alpha =\frac{\lambda _{2}-h_{11}^{0}}{\lambda _{2}-\lambda _{1}}%
\end{array}%
\right.
\end{equation}%
where the $\lambda _{j},h_{jk}^{0}$ can depend on frequency $\omega /2\pi $
but are independent of $z$\ and%
\begin{equation}
\left\{ 
\begin{array}{c}
\lambda _{1}=\frac{1}{2}\left( h_{11}^{0}+h_{22}^{0}+\sqrt{\rho }e^{i\theta
/2}\right) \\ 
\lambda _{2}=\frac{1}{2}\left( h_{11}^{0}+h_{22}^{0}-\sqrt{\rho }e^{i\theta
/2}\right) \\ 
\Delta =\left( h_{11}^{0}-h_{22}^{0}\right) ^{2}+4h_{12}^{0}h_{21}^{0}=\rho
e^{i\theta }.%
\end{array}%
\right.
\end{equation}%
The eigenvalues have real parts strictly negative (to fulfill the condition $%
\left( 12\right) ).$

\textbf{Case\ 2: }$\lambda _{1}=\lambda _{2}$

The solutions are given by the equalities%
\begin{equation}
\left\{ 
\begin{array}{l}
H_{11}^{z}=\left( az+1\right) e^{\lambda z},\text{ \ \ }%
H_{12}^{z}=h_{12}^{0}e^{\lambda z} \\ 
H_{22}^{z}=\left( -az+1\right) e^{\lambda z},\text{ \ \ }%
H_{21}^{z}=h_{21}^{0}e^{\lambda z} \\ 
a=\frac{h_{11}^{0}-h_{22}^{0}}{2},\text{ \ \ \ \ }\lambda =\frac{%
h_{11}^{0}+h_{22}^{0}}{2}.%
\end{array}%
\right.
\end{equation}

The case $h_{11}^{0}=h_{22}^{0}\neq 0,h_{12}^{0}=h_{21}^{0}=0$ leads to the
usual B.L law. These equalities are verified in any system of coordinates.
Components evolve independently, with same attenuation and celerity. This
corresponds to a medium with all possible properties of symmetry.

In all cases, the real part of eigenvalues different from 0 have to be
negative for passive media which weaken waves. Moreover, the solutions are
only matched to equations $\left( 11\right) ,\left( 12\right) ,\left(
13\right) .$

\section{Examples}

In examples, we assume that the parameters $h_{jk}^{0}\left( \omega \right) $
are constant on spectral supports of inputs \textbf{E}$_{x}^{0},$\textbf{E}$%
_{y}^{0}$.

\subsection{Example 1}

Let assume that 
\begin{equation*}
h_{11}^{0}\neq h_{22}^{0}\text{ and }h_{12}^{0}=h_{21}^{0}=0.
\end{equation*}%
We are in the case 1 with 
\begin{equation*}
H_{11}^{z}=e^{h_{11}^{0}z},H_{22}^{z}=e^{h_{22}^{0}z},H_{12}^{z}=H_{21}^{z}=0.
\end{equation*}%
This means that a beam $e^{i\omega t}$ polarized on Ox is transmitted with
weakening exp$\left[ \mathcal{R}\left[ h_{11}^{0}\right] z\right] $ and
delay $\mathcal{I}\left[ h_{11}^{0}\right] z/\omega .$ If polarized along
Oy, the weakening is exp$\left[ \mathcal{R}\left[ h_{22}^{0}\right] z\right] 
$ and the delay is $\mathcal{I}\left[ h_{22}^{0}\right] z/\omega .$ When $%
\mathcal{R}\left[ h_{11}^{0}\right] /\mathcal{R}\left[ h_{22}^{0}\right] \ll
1,$ the first component disappears before the second component,
independently of the values of $\mathcal{I}\left[ h_{11}^{0}\right] $ and $%
\mathcal{I}\left[ h_{22}^{0}\right] $ which define the refraction indices of
the medium. Then we can give to $h_{11}^{0},h_{22}^{0}$ values fitted to a
dichroic material. More generally when%
\begin{equation*}
E_{x}^{z}\left( t\right) =e^{h_{11}^{0}z}E_{x}^{0}\left( t\right) ,\
E_{y}^{z}\left( t\right) =e^{h_{22}^{0}z}E_{y}^{0}\left( t\right)
\end{equation*}%
both components evolve independently and the usual B.L law is verified for
each component. The power $P_{z}$ at $z$ becomes (we have assumed that the $%
h_{jk}^{0}$ are constant with respect to frequency)%
\begin{equation*}
P_{z}=e^{2z\mathcal{R}\left[ h_{11}^{0}\right] }\sigma _{x}^{2}+e^{2z%
\mathcal{R}\left[ h_{22}^{0}\right] }\sigma _{y}^{2}
\end{equation*}%
where $\sigma _{x}^{2}=$E$\left[ \left\vert E_{x}^{0}\left( t\right)
\right\vert ^{2}\right] ,\sigma _{y}^{2}=$E$\left[ \left\vert
E_{y}^{0}\left( t\right) \right\vert ^{2}\right] .$ We retrieve the usual
result when $\mathcal{R}\left[ h_{11}^{0}\right] =\mathcal{R}\left[
h_{22}^{0}\right] .$ In other cases usual B.L law is no longer valid because
the two terms in $P_{z}$ have different behaviors.

\subsection{Example 2}

Let assume that we are in the case 1 (two distinct eigenvalues $\lambda
_{1},\lambda _{2}$ different from 0) and that we deal with a monochromatic
wave polarized along Ox: 
\begin{equation*}
E_{x}^{0}\left( t\right) =e^{i\omega _{0}t},\text{ \ }E_{y}^{0}\left(
t\right) =0.
\end{equation*}%
By $\left( 8\right) $ we have%
\begin{equation*}
\mathbf{E}_{x}^{z}=\mathcal{H}_{11}^{z}\left[ \mathbf{E}_{x}^{0}\right] ,%
\text{ \ }\mathbf{E}_{y}^{z}=\mathcal{H}_{21}^{z}\left[ \mathbf{E}_{x}^{0}%
\right] .
\end{equation*}%
\textbf{E}$^{0}$ is transformed in \textbf{E}$^{z}$ defined by 
\begin{equation*}
\left\{ 
\begin{array}{l}
E_{x}^{z}\left( t\right) =\left[ \frac{\lambda _{2}-h_{11}^{0}}{\lambda
_{2}-\lambda _{1}}e^{\lambda _{1}z}+\frac{h_{11}^{0}-\lambda _{1}}{\lambda
_{2}-\lambda _{1}}e^{\lambda _{2}z}\right] e^{i\omega _{0}t}\text{ \ } \\ 
E_{y}^{z}\left( t\right) =\frac{-h_{21}^{0}}{\lambda _{2}-\lambda _{1}}%
\left( e^{\lambda _{1}z}-e^{\lambda _{2}z}\right) e^{i\omega _{0}t}%
\end{array}%
\right.
\end{equation*}%
where parameters can be complex$.$ At $z,$ both components are weakened and
delayed through two terms functions of $z$ ($e^{\lambda _{1}z}$ and $%
e^{\lambda _{2}z})\ $and not one as in the usual B.L law.

1) When $h_{21}^{0}=0$ we have%
\begin{equation*}
E_{x}^{z}\left( t\right) =e^{h_{11}^{0}z+i\omega _{0}t},\text{ \ \ }%
E_{y}^{z}\left( t\right) =0.
\end{equation*}%
\textbf{E}$^{z}$ is polarized along Ox with $\mathcal{R}\left[ h_{11}^{0}%
\right] $ and $\mathcal{I}\left[ h_{11}^{0}\right] $ as parameters of
weakening and of delay. $P_{z}$ $=e^{2z\mathcal{R}\left[ h_{11}^{0}\right] }$
has the shape $\left( 5\right) $ of the usual B.L law.

2) Except when $h_{21}^{0}=0,$ the monochromatic wave \textbf{E}$^{z}$ is no
longer polarized. Assume that $h_{21}^{0}\neq 0,h_{12}^{0}=0$ (which implies 
$h_{11}^{0}\neq 0,h_{22}^{0}\neq 0,h_{11}^{0}\neq h_{22}^{0}).$ From $\left(
14\right) $%
\begin{equation*}
\left\{ 
\begin{array}{l}
E_{x}^{z}\left( t\right) =e^{h_{11}^{0}z+i\omega _{0}t}\text{ \ } \\ 
E_{y}^{z}\left( t\right) =\frac{h_{21}^{0}}{h_{11}^{0}-h_{22}^{0}}\left(
e^{h_{11}^{0}z}-e^{h_{22}^{0}z}\right) e^{i\omega _{0}t}.%
\end{array}%
\right.
\end{equation*}%
Even when $\mathcal{R}\left[ h_{11}^{0}\right] =\mathcal{R}\left[ h_{22}^{0}%
\right] $ it is impossible to have $P_{z}$ like $\left( 5\right) .$
Consequently we understand using symmetries that usual BL law can be true
for a particular polarization, and untrue for other polarizations.

\subsection{Example 3}

1) The case 
\begin{equation}
h_{12}^{0}=-h_{21}^{0},\text{ \ }h_{11}^{0}=h_{22}^{0}\neq 0
\end{equation}%
is particularly interesting. We have%
\begin{equation*}
\lambda _{1}=h_{11}^{0}+ih_{12}^{0},\text{ \ }\lambda
_{2}=h_{11}^{0}-ih_{12}^{0}
\end{equation*}%
and (14) becomes%
\begin{equation*}
\left\{ 
\begin{array}{c}
H_{11}^{z}=H_{22}^{z}=\frac{1}{2}\left( e^{\lambda _{1}z}+e^{\lambda
_{2}z}\right) \\ 
H_{12}^{z}=-H_{21}^{z}=-\frac{i}{2}\left( e^{\lambda _{1}z}-e^{\lambda
_{2}z}\right) .%
\end{array}%
\right.
\end{equation*}%
Now we assume that the electric field \textbf{E}$^{0}$ is monochromatic and
polarized at angle $\phi $ with respect to Ox 
\begin{equation*}
\left\{ 
\begin{array}{c}
E_{x}^{0}\left( t\right) =e^{i\omega _{0}t}\cos \phi \\ 
E_{y}^{0}\left( t\right) =e^{i\omega _{0}t}\sin \phi .%
\end{array}%
\right.
\end{equation*}%
From ($8)$ and after elementary algebra%
\begin{equation}
\left\{ 
\begin{array}{c}
E_{x}^{z}\left( t\right) =e^{i\omega _{0}t+h_{11}^{0}z}\cos \left( \phi
-zh_{12}^{0}\right) \\ 
E_{y}^{z}\left( t\right) =e^{i\omega _{0}t+h_{11}^{0}z}\sin \left( \phi
-zh_{12}^{0}\right) .%
\end{array}%
\right.
\end{equation}%
For real $h_{12}^{0},$ the electric field \textbf{E}$^{z}$ at $z$ is
polarized at the angle $\left( \phi -zh_{12}^{0}\right) .$ Then a
monochromatic wave at the frequency $\omega _{0}/2\pi $ is rotated by the
angle $-zh_{12}^{0}$ and attenuated by $\exp \left( z\mathcal{R}\left[
h_{11}^{0}\right] \right) $. Moreover $-\frac{z}{\omega _{0}}\mathcal{I}%
\left[ h_{11}^{0}\right] $ is the time spent by the wave between O and $z$
which shows that the medium refraction index $n\left( \omega _{0}\right) $
is equal to ($c$ is the light velocity in vacuum)%
\begin{equation*}
n\left( \omega _{0}\right) =-\frac{c}{\omega _{0}}\mathcal{I}\left[
h_{11}^{0}\right] .
\end{equation*}%
The rotation angle is independent of the polarization angle $\phi .$ Because 
$P_{z}=e^{2z\mathcal{R}\left[ h_{11}^{0}\right] }$ usual B.L law is obeyed
for the amplitude and the power, but not for components.

2) When $h_{12}^{0}$ is not real $\left( 18\right) $ is developed in%
\begin{equation}
\left\{ 
\begin{array}{c}
E_{x}^{z}\left( t\right) =e^{i\omega _{0}t+h_{11}^{0}z}(\cos a\cosh b+i\sin
a\sinh b) \\ 
E_{y}^{z}\left( t\right) =e^{i\omega _{0}t+h_{11}^{0}z}(\sin a\cosh b-i\cos
a\sinh b) \\ 
a=\phi -z\mathcal{R}\left[ h_{12}^{0}\right] ,\text{ \ }b=z\mathcal{I}\left[
h_{12}^{0}\right] .%
\end{array}%
\right.
\end{equation}%
Consequently \textbf{E}$^{z}$ has two components polarized in orthogonal
directions of angles $a$ and $\left( a+\frac{\pi }{2}\right) $ with respect
to Ox and of "height" $B_{1}^{z}$ and $B_{2}^{z}$ such that%
\begin{equation}
\left\{ 
\begin{array}{c}
B_{1}^{z}=e^{\mathcal{R}\left[ h_{11}^{0}\right] z}\cosh \left( z\mathcal{I}%
\left[ h_{12}^{0}\right] \right) \\ 
B_{2}^{z}=-ie^{\mathcal{R}\left[ h_{11}^{0}\right] z}\sinh \left( z\mathcal{I%
}\left[ h_{12}^{0}\right] \right) \\ 
\left\vert \frac{B_{2}^{z}}{B_{1}^{z}}\right\vert =\tanh \left\vert z%
\mathcal{I}\left[ h_{12}^{0}\right] \right\vert .%
\end{array}%
\right.
\end{equation}%
$\left\vert \frac{B_{2}^{z}}{B_{1}^{z}}\right\vert $ increases from 0 to 1
when $z$ increases from 0 to $\infty .$ The main component is rotated by $-z%
\mathcal{R}\left[ h_{12}^{0}\right] $ and attenuated by $\exp \left( z%
\mathcal{R}\left[ h_{11}^{0}\right] \right) \cosh b$. In the same time a
second component appears which is orthogonal to the first component. The
ratio of "heights" increases with $z$ and tends towards 1. The power $P_{z}$
at $z$ is given taking into account othogonality of components%
\begin{equation*}
P_{z}=e^{2z\mathcal{R}\left[ h_{11}^{0}\right] }\cosh \left( 2z\mathcal{I}%
\left[ h_{12}^{0}\right] \right)
\end{equation*}%
which shows that the usual B.L law is not valid (except in the case of real $%
h_{12}^{0}$ which leads to $P_{z}=e^{2z\mathcal{R}\left[ h_{11}^{0}\right]
}) $. The power of each component $\mathbf{E}_{x}^{z},\mathbf{E}_{y}^{z}$
does not follow the usual B.L law.~For instance, for the power of the
component $\mathbf{E}_{x}^{z}$ 
\begin{equation*}
P_{x}^{z}=e^{2z\mathcal{R}\left[ h_{11}^{0}\right] }\left( \cos ^{2}\left(
\phi -z\mathcal{R}\left[ h_{12}^{0}\right] \right) +\sinh ^{2}\left( z%
\mathcal{I}\left[ h_{12}^{0}\right] \right) \right) .
\end{equation*}%
The same remark is true for components in the directions $a$ and $\left( a+%
\frac{\pi }{2}\right) $ with respect to Ox.

\subsection{Example 4}

We assume that \textbf{E}$^{0}$ is a quasi-monochromatic beam (see section 1)%
\begin{equation*}
E_{x}^{0}\left( t\right) =e^{i\omega _{0}t}A_{x}^{0}\left( t\right) ,\text{
\ }E_{y}^{0}\left( t\right) =e^{i\omega _{0}t}A_{y}^{0}\left( t\right) .
\end{equation*}%
The $h_{jk}^{0}\left( \omega \right) $ are constant with respect of $\omega $
on the beam spectrum. If the conditions $\left( 17\right) $ are fulfilled,
the components \textbf{E}$_{x}^{0},$\textbf{E}$_{y}^{0}$ are split by the
medium in two orthogonal parts. With respect to $Ox^{\prime }y^{\prime }$
defined by%
\begin{equation*}
\left( Ox,Ox^{\prime }\right) =\left( Oy,Oy^{\prime }\right) =-z\mathcal{R}%
\left[ h_{12}^{0}\right]
\end{equation*}%
the beam at $z$ verifies using $\left( 19\right) $ and $\left( 20\right) $%
\begin{equation*}
\left\{ 
\begin{array}{c}
E_{x^{\prime }}^{z}\left( t\right) =e^{\mathcal{R}\left[ h_{11}^{0}\right] z}%
\left[ A_{x}^{0}\left( t\right) \cosh \left( z\mathcal{I}\left[ h_{12}^{0}%
\right] \right) +A_{y}^{0}\left( t\right) \sinh \left( z\mathcal{I}\left[
h_{12}^{0}\right] \right) \right] \\ 
E_{y^{\prime }}^{z}\left( t\right) =e^{\mathcal{R}\left[ h_{11}^{0}\right] z}%
\left[ A_{y}^{0}\left( t\right) \cosh \left( z\mathcal{I}\left[ h_{12}^{0}%
\right] \right) -A_{x}^{0}\left( t\right) \sinh \left( z\mathcal{I}\left[
h_{12}^{0}\right] \right) \right] .%
\end{array}%
\right.
\end{equation*}%
We deduce the power $P_{z}$ defined by $\left( 6\right) $%
\begin{equation*}
\left\{ 
\begin{array}{c}
P_{z}=e^{2\mathcal{R}\left[ h_{11}^{0}\right] z}\left[ P_{0}\cosh \left( 2z%
\mathcal{I}\left[ h_{12}^{0}\right] \right) +\theta \sinh \left( 2z\mathcal{I%
}\left[ h_{12}^{0}\right] \right) \right] \\ 
\theta =2\mathcal{I}\left\{ \text{E}\left[ A_{x}^{0}\left( t\right)
A_{y}^{0\ast }\left( t\right) \right] \right\} .%
\end{array}%
\right.
\end{equation*}%
Whatever the polarization state of the beam at $z=0,$\ the usual B.L law is
verified if and only if $h_{12}^{0}$ is real, i.e when the effect of the
medium is a rotation (added to a weakening). Whatever $h_{12}^{0},$ we have $%
\theta =0$ for instance when \textbf{E}$^{0}$ is polarized (E$\left[ ..%
\right] $ is real) or unpolarized (\textbf{E}$_{x}^{0}$ and \textbf{E}$%
_{y}^{0}$ are uncorrelated).

\subsection{Example 5}

We assume that \textbf{E}$^{0}$ is a quasi-monochromatic beam like $\left(
7\right) $ and%
\begin{equation*}
h_{11}^{0}=h_{22}^{0},\text{ \ }h_{12}^{0}=h_{21}^{0}
\end{equation*}%
i.e transfers between the coordinates are symmetric. From $\left( 8\right) $
we have%
\begin{equation*}
\left\{ 
\begin{array}{c}
H_{11}^{z}=H_{22}^{z}=e^{h_{11}^{0}z}\cosh h_{12}^{0}z \\ 
H_{12}^{z}=H_{21}^{z}=e^{h_{11}^{0}z}\sinh h_{12}^{0}z.%
\end{array}%
\right.
\end{equation*}%
However we remark that these relations are not maintained in other reference
systems. For instance, in $Ox^{\prime }y^{\prime }$ with $\left(
Ox,Ox^{\prime }\right) =\phi $ we have (the $k_{ij}^{0}$ are the new
parameters)%
\begin{equation*}
\left\{ 
\begin{array}{c}
k_{11}^{0}=h_{11}^{0}+h_{12}^{0}\sin 2\phi \\ 
k_{22}^{0}=h_{11}^{0}-h_{12}^{0}\sin 2\phi \\ 
k_{12}^{0}=k_{21}^{0}=h_{12}^{0}\cos 2\phi .%
\end{array}%
\right.
\end{equation*}%
Elementary algebra leads to%
\begin{equation*}
\left\{ 
\begin{array}{c}
P_{z}=e^{2\mathcal{R}\left[ h_{11}^{0}\right] z}\left[ P_{0}\cosh \left( 2z%
\mathcal{R}\left[ h_{12}^{0}\right] \right) +\theta ^{\prime }\sinh \left( 2z%
\mathcal{R}\left[ h_{12}^{0}\right] \right) \right] \\ 
\theta ^{\prime }=2\mathcal{R}\left\{ \text{E}\left[ A_{x}^{0}\left(
t\right) A_{y}^{0\ast }\left( t\right) \right] \right\}%
\end{array}%
\right.
\end{equation*}%
which proves that the usual B.L law is not true apart from particular cases.

\section{Conclusion}

The Beer-Lambert law (actually due to P. Bouguer around 1729) was firstly
used to measure the concentration of solutions . It addresses the problem of
concentration measurement of some kind of molecules in a liquid. In equation
1, we have 
\begin{equation}
\alpha =k\left( \omega \right) a
\end{equation}%
where $a$ is the concentration and $k$ $\left( \omega \right) $ is a
wavelength-dependent absorptivity coefficient. $k$ is deduced from a
measurement of the attenuation for a known value of $a.$ The property of
linearity with respect to the distance is due to Lambert \ and the linearity
with respect to the concentration of absorbing species in the material was
highlighted by Beer. The BL law intervenes in wave propagation to explain
together the attenuation and the dispersion whatever the crossed medium.

A version of the Beer-Lambert law addresses the power as a function of the
medium thickness, whatever the nature of the wave, acoustic or
electromagnetic. The result has the form $A\left( z\right) =A\left( 0\right)
e^{-\alpha z}$ where $\alpha $ is a function of the medium and of the
frequency $\omega /2\pi $. Very often $\alpha $ is a power function of $%
\omega .$ Its estimation has numerous applications in medecine \cite{Lewi}, 
\cite{Park}. Also, chemistry uses the measurements of $\alpha $ because it
is a linear function of the number of particles imbedded in the medium. The
B.L law can be wrong for instance when multipaths in fibres or in case of
too strong transmitted powers \cite{Abit}, \cite{Alle}. Moreover the medium
and the devices can be sensitive to polarization state. For instance the B.L
law can be untrue in the case of birefringence for the medium and when
devices select only one component of the field. However a generalization is
possible studying separately both components of the electric field which
defines an electromagnetic beam.

The B.L law is easily proved modelling a medium thickness as a linear
invariant filter (LIF) where input and output show the evolution of the
quantity of interest (for instance an amplitude or a power). In this paper
we study the evolution of two quantities, the components of the electric
field of an electromagnetic beam. To take into account interactions between
components, a piece of medium is modelled by four LIF. We assume that the
crossings of two successive medium pieces are independent events. This
hypothesis suffices to determine the shape of the LIF complex gains. They
are defined by four parameters $h_{11}^{0},$ $h_{12}^{0}$ $,h_{21}^{0}$ $%
,h_{22}^{0}$ which depend on the medium and which may depend on the
frequency. This model generalizes the Jones matrices used in the
deterministic monochromatic beam to stationary random beams. As soon
explained, Jones papers do not contain comments about the BL law. Examples
of section 3 show that the set of parameters can be fitted to realistic
situations.

\end{document}